\documentclass[12pt,preprint]{aastex}

\def\arcsec{\char'175 }

\def\etal{et~al.\ }

\def\hub{\ifmmode H_\circ\else H$_\circ$\fi}

\shorttitle{Coud\'e Feed Spectral Library}
\shortauthors{Valdes, Gupta, Rose, Singh, \& Bell}

\begin{document}

\title{The Indo-U.S. Library of Coud\'e Feed Stellar Spectra}
\author{Francisco Valdes}
\affil{National Optical Astronomy Observatory, P.O. Box 26732, Tucson, AZ 85726}

\and

\author{Ranjan Gupta}
\affil{IUCAA, Post Bag 4, Ganeshkhind, Pune 411 007, India}

\and

\author{James A. Rose}
\affil{Department of Physics and Astronomy, CB \#3255, University of North 
Carolina, Chapel Hill, NC 27599}

\and

\author{Harinder P. Singh}
\affil{Department of Physics \& Astrophysics, University of Delhi, Delhi 
110 007, India}

\and

\author{David J. Bell}
\affil{National Optical Astronomy Observatory, P.O. Box 26732, Tucson, AZ 85726}
\email{fvaldes@noao.edu, rag@iucaa.ernet.in, jim@physics.unc.edu, hpsingh@physics.du.ac.in, dbell@noao.edu}
      
\begin{abstract}
We have obtained spectra for 1273 stars using the 0.9m Coud\'e Feed telescope
at Kitt Peak National Observatory.  This telescope feeds the coud\'e
spectrograph of the 2.1m telescope.  The spectra have been obtained with
the \#5 camera of the coud\'e spectrograph and a Loral 3K X 1K CCD.
Two gratings have been used to provide spectral coverage from 3460~\AA \ to
9464~\AA, at a resolution of $\sim$1\AA \ FWHM and at an
original dispersion of 0.44~\AA/pixel.  For 885 stars we have complete spectra
over the entire 3460~\AA \ to 9464~\AA \ wavelength region (neglecting small
gaps of $<$ 50 \AA), and partial spectral
coverage for the remaining stars.  The 1273 stars have been selected to
provide broad coverage of the atmospheric parameters T$_{eff}$, log g,
and [Fe/H], as well as spectral type.  The goal of the project is
to provide a comprehensive library of stellar spectra for use in the
automated classification of stellar and galaxy spectra and in galaxy
population synthesis.  In this paper we discuss the characteristics of
the spectral library, viz., details of the observations, data reduction
procedures, and selection of stars.  We also present a few illustrations of
the quality and information available in the spectra.  The first version of
the complete spectral library is now publicly available from the National
Optical Astronomy Observatory (NOAO) via FTP and HTTP.
\end{abstract}


\section{Introduction}

The need for a comprehensive database of digital stellar spectra covering a 
large range in T$_{eff}$, log g, and [Fe/H] at moderate (1-2 \AA \ FWHM) 
spectral resolution has increased considerably in the past decade.  The uses
of such a spectral library are wide ranging.  To begin with, synthetic stellar
spectra generated from model atmospheres now incorporate such extensive line
lists (e.g., Kurucz 1993, 1994; Bell \& Gustaffson 1978, 1989; Tripicco \& Bell
1992, 1995) that the synthetic spectra
can be compared to empirical spectra at increasingly high spectral resolution.  
Once the basic accuracy of the synthetic spectra has been established, 
the synthetic spectrum technique can then be utilized to explore areas of 
atmospheric parameter space that are not adequately represented in Solar 
Neighborhood stars.  For example, recent results from galaxy population
synthesis studies (e.g., Kuntschner \etal 2002; Terlevich \& Forbes 2002; 
Trager \etal 2000 and references therein) indicate that the integrated spectra 
of most 
early-type galaxies are dominated by metal-rich stars with non-solar abundance 
ratios that are not found in the Solar Neighborhood. 
Second, with the advent of large databases of
both stellar and galaxy spectra, generated by multi-fiber and 
multi-aperture spectrographs, methods for the automated parametrization of 
spectra in a fast and reliable manner are becoming essential.  Indeed, various 
researchers have been exploring the application of Artificial Neural Networks 
to the automated parametrization of stellar spectra (e.g., Singh, Bailer-Jones,
\& Gupta 2002).  In addition, 
Katz \etal (1998) have developed the TGMET software, which establishes the best 
match between a target spectrum and a library of reference spectra.  For these
techniques to be applied, a comprehensive set of reference spectra with known
atmospheric parameters is essential.  Finally, in the field of spectral
synthesis of the integrated light of galaxies, the need to resolve individual
features in galaxy spectra, to effectively resolve the problems of 
non-uniqueness in extracting mean age and metallicity information from 
integrated spectra, has become increasingly evident (Worthey 1994; Vazdekis \&
Arimoto 1999).  To carry out
the modeling of the composite spectra of galaxies of different ages and 
metallicities requires a spectral database that covers all areas in the HR
diagram sampled by theoretical isochrones.

Despite the clear necessity for a comprehensive spectral library, there
is surprisingly little existing material to select from.  In fact, until 
recently the existing spectral libraries have primarily consisted of low
resolution (5 - 20 \AA \ FWHM) spectrophotometry of typically $\sim$100 stars,
covering a range in spectral type, but largely restricted to solar chemical
composition (Pickles 1998 and references therein).  This traditional emphasis on accurate 
spectrophotometry with
broad wavelength coverage at low resolution has likely reflected both the low
number of pixels in spectroscopic detectors until recently, coupled with the
low resolution approach to spectral synthesis studies of stars and galaxies 
that supplied the chief driver for the libraries.  However, the advent of
large format CCD detectors in spectroscopy, along with the above-mentioned 
trends toward higher resolution modeling of the spectra of stars and of 
galaxies in integrated light, is now leading to the production of several 
extensive libraries of stellar spectra at higher spectral resolution. 
In particular, Jones (1999, see also Leitherer \etal 1996) completed a spectral
database of 684 stars at 2 \AA \ resolution (FWHM) using the Coud\'e Feed
telescope at the Kitt Peak National Observatory (KPNO).  Due to the small format
of the existing CCD detector at that time, coverage is restricted to two 
$\sim$700 \AA \ spectral regions.  More recently, Prugniel \& Soubiran (2001)
published a spectral library, based on the ELODIE echelle spectrograph at the
Observatoire de Haute-Provence, that covers the wavelength interval 
$\lambda$$\lambda$4100-6800 \AA \ at a resolution of R=42000 for 708 stars that
cover a large range in atmospheric parameters.  Finally, Cenarro \etal (2001)
have observed 706 stars in the
wavelength region $\lambda$$\lambda$8350-9020~\AA \ at 1.5 \AA \ resolution, to
characterize the behavior of the near-IR Ca II triplet for galaxy population
synthesis.

In this paper we describe a new database of stellar spectra, encompassing more
than 1000 stars, and covering the spectral region $\lambda$$\lambda$3400--9500
\AA \ at a resolution of $\sim$1 \AA \ FWHM.  The primary emphasis of this new
spectral library is on broad wavelength coverage, particularly well into the 
blue, at a resolution sufficient to resolve numerous diagnostic spectral
features that can be used in the automated parametrization of spectra and in
the population synthesis of galaxies.  We have also emphasized
coverage in atmospheric parameter space, particularly at lower metallicity.
The purpose of this paper is to provide a guide to the new spectral library,
which is now publicly available from the National Optical Astronomy Observatory
(NOAO) via ftp and http.  The paper presents technical information about the 
observations and the sample of stars observed to make the database
readily useful to the astronomical 
community.  In \S2 we describe the observational procedures used in creating the
library, and in \S3 we describe how the sample of stars was selected.  In
\S4 the data reduction procedures are described.  In \S5 we describe the actual
format of the archived spectral database as well as the tables which summarize
the key information about the stars.  Finally, in \S6 we show examples of the
stellar spectra, some of which are meant to illustrate the challenges facing
the automated parametrization of stellar spectra.

\section{Observations}

All observations for the spectral library have been obtained with the Coud\'e Feed
telescope at KPNO.  The coud\'e feed is an auxiliary telescope (a 60'' flat 
reflects the sky onto a 36'' parabolic mirror which directs a converging beam
into the 2.1-m telescope coud\'e optical train) which feeds a stellar image onto the
slit of the coud\'e spectrograph of the 2.1-m telescope.  Two spectrograph
combinations were optimized for the red and blue portions of the spectra.
For the blue the KPC10A grating (316 lines/mm) and blue corrector were used
while for the red the B\&L181 grating (316 lines/mm) and red corrector were
used.  In common were the \#5 camera, the small collimator, and the
3K x 1K Loral CCD detector (F3KB).  These combinations produce a fairly
uniform dispersion of 0.44 \AA/pixel yielding $\sim$1300 \AA~coverage
along the 3K dimension of the the CCD.  The orientation of the CCD
is such that spectra are dispersed as nearly parallel to the serial
direction as possible.  Any CTE problems then only affect the spectral
profile and not the spectral resolution.

Consequently, the entire spectral region
$\lambda$$\lambda$3400--9500 \AA \ can be covered with 5 grating settings, 
albeit with only a little overlap between settings.  With a slit width of
200 $\micron$, or 1.4\arcsec, the focus is typically $\sim$3 pixels FWHM,
or $\sim$1.2 \AA.  A slit length of 9.77 mm, or 70.6\arcsec, provides 100
pixels for source and sky.
While we focussed the spectrograph to obtain roughly constant
focus across the spectrum, inevitably there are small focus variations across
specific 1300 \AA \ grating settings, as well as small variations from night to 
night, despite the relatively stable environment in the Coud\'e Feed room.  
However, we anticipate that maximum use of the spectral database will be at
typically 2 \AA \ resolution, hence the slight focus variations along the 
spectrum will not be a major problem.

Exposure times at the various grating settings were adjusted to target a
S/N ratio of better than 100:1 per 0.44 \AA /pixel.
However, when the seeing was poor, in regions of low flux and detector
DQE, and for the faintest stars (B$>$10) such a high S/N ratio was not
achievable.  For the faintest stars we simply used a maximum exposure
time of 20 minutes.  The practical limit for the library is B=12.  

Given that the coud\'e slit cannot be easily rotated to the parallactic
angle or tracked with the field rotation inherent in the feed arrangement,
and that
a narrow (1.4\arcsec) slit is required to maintain spectral resolution, 
obtaining accurate spectrophotometry for the survey was not feasible.  The problem is further
exacerbated by the fact that the 60'' feed mirror is 
particularly subject to wind shake.  We typically acquire observations of
three spectrophotometric standard stars during the night, to achieve
approximate spectrophotometry.  Radial velocity accuracy is also not an
important consideration.  
The procedure while observing
is to take a series of arc lamp exposures before twilight to build up a master
arc exposure, then shorter exposures are obtained approximately on an hourly
basis during the night.  Given that the
spectrograph is off the telescope in a thermally stable environment, the zero
point shift is independent of the RA and DEC of the observation.

Altogether 6917 individual spectra of 1273 stars have been
obtained during the time period from June 1995 through April 2003.
The project has been a part-time effort over many years with changes in
personnel and goals.  As a consequence, on several
observing runs, grating settings were used that do not provide full overlap
with adjacent settings, leaving a considerable number of final spectra with 
small ($<$50 \AA) gaps.  The project has involved the participation of many
present and former undergraduate and graduate students at the University of
North Carolina, which has greatly aided in its completion.

\section{Selection of Stars}

The primary emphasis for the spectral database is on obtaining stellar spectra
covering a wide range in T$_{eff}$, log g, and [Fe/H].  Naturally, it is
important to select stars whose atmospheric parameters are well-determined from
a fundamental (i.e., high dispersion) analysis,
and preferably part of a large homogeneous sample.  A prototypical example of
a uniform sample of high-quality data is the Edvardsson \etal (1993) study, in
which a high dispersion abundance analysis has been conducted for 189 F and 
early G stars.  Our first priority has been on working with samples such as the
Edvardsson \etal (1993) one, in which a number of stars have been
subjected to a well-defined analysis that includes elemental abundance ratios,
and thus are placed on a highly reliable 
relative scale.  Furthermore, if subsequent work indicates that the effective
temperature scale used in the analysis requires revision, the prescription is 
available to readily redefine both T$_{eff}$ and [Fe/H] values for the entire
sample.  
Abundance studies that fit these criteria and from which we have observed stars
for our spectral database are Castro \etal (1997), Chen \etal (2000), Edvardsson 
\etal (1993), Feltzing \& Gustafsson (1998), Luck \& Challener (1995), Nissen \&
Schuster (1997), Tautvaisiene (1997), and Tautvaisiene \etal (2001).  The
individual samples are further summarized in \S5.2.

A second type of sample regards large compilations,
homogeneous in quality, but for which the atmospheric parameter information is
less extensive and/or fundamental than for the samples mentioned above.  An example is the Dickow \etal (1970)
multi-passband photometric survey of giants, for which atmospheric parameters
have been extracted by Hansen \& Kjaergaard (1971). A second example is
the work of Brown \etal
(1989), also for giants, for which the T$_{eff}$ information has been taken
from the DDO photometry of McClure \& Forrester (1982), and [Fe/H] extracted 
from high-resolution spectroscopy over a very limited wavelength region (thus
precluding a detailed elemental abundance analysis).  Similarly, the atmospheric
parameters for many of the metal-poor stars in the database are taken from
Carney \etal (1994), where results are based on the combination of a photometric
effective temperature and [Fe/H] from low S/N ratio high dispersion spectra in
the Mg I triplet region.  Another example is the
previously mentioned Prugniel \& Soubiran (2001) survey of 709 stars.  While 
high dispersion spectra have been obtained for all stars, the atmospheric
parameters are indirectly obtained from comparison of the star's spectrum with
various template spectra.  Further descriptions are given in \S5.2.

A third category of stars in our database are those for which atmospheric
parameters have been obtained for just one or a few stars, hence the 
literature values are subject to the additional systematic uncertainties 
associated with such targeted studies.  For the most part, stars in this
category have been included because they help to define regions of
atmospheric parameter space which are otherwise sparsely represented in our
database, such as F and G supergiants.  The atmospheric parameter data for the
majority of these stars have been taken from the monumental compilation of
Cayrel de Strobel, Soubiran, \& Ralite (2001) and, in some cases, from
Cayrel de Strobel \etal (1997).

Finally, there are a number of stars in the database for which well-determined
atmospheric parameters are lacking.  These are primarily stars used to cover
areas of parameter space which are under-represented. For example, we have
observed a number of stars with supergiant spectral classifications, but without
atmospheric parameter determinations, in order to enhance our sample of 
low-gravity stars.  Likewise, we have observed a number of stars with
early-type spectral classifications, but no additional information, to have
some coverage of hot stars.  In 
addition, we have observed a number of stars from Rose (1985) that are found
to be strong-lined G dwarfs and are believed to be metal-rich, similar in fact 
to stars in the Cayrel de Strobe, Soubiran, \& Ralite (2001) compendium that have 
[Fe/H]$\sim$+0.3.  However, no fundamental abundance analysis has been carried
out for these stars. For these latter stars we have assigned a log g of 4.0, 
and [Fe/H]=+0.3, but these parameters must be considered highly
provisional at this point.  In addition, many of the metal-poor dwarfs studied
by Carney \etal (1994) do not have a determination of log g.  We have set 
log g = 4.0 for these stars. 

In assigning atmospheric parameters to the stars in our sample, we used the
following algorithm.  We first checked both the Cayrel de Strobel \etal (1997)
catalog of [Fe/H] determinations for all stars and the Cayrel de Strobel, Soubiran, \& Ralite 
(2001) catalog of [Fe/H] determinations for F,G, and K stars.  If values are
given for the star in Cayrel de Strobel, Soubiran, \& Ralite (2001), we used those over the
1997 catalog.  In the Cayrel de Strobel \etal catalogs the atmospheric 
parameters from all determinations for a given
star are listed in chronological order.  We first check to see whether the
star has values reported from Edvardsson \etal (1993), Luck \& Challener (1995),
Castro \etal (1997), Feltzing \& Gustafsson (1998), or Chen \etal (2000).  If
any of these sources are available, we default to those values, since they 
represent homogeneous compendiums.  If the star is contained in more than one
of these references, we select in the order given above, i.e., Edvardsson
\etal (1993) has highest priority.  If the star has not been covered by any
of the above five references, then we default to the most recent atmospheric
parameters listed in Cayrel de Strobel, Soubiran, \& Ralite (2001) which have complete information
for all three parameters and also do not have a ``quality'' flag indicating 
lower quality data.  

Overall, we have obtained good coverage over a large region of parameter space
in T$_{eff}$, log g, and [Fe/H].  However, there are some areas with relatively
low representation.  First, it has been difficult to find an 
adequate sample of moderately metal-poor giants with -0.6$\geq$[Fe/H]$\geq$-1.2.
Due to the modest aperture of the Coud\'e Feed system and the goal of a large 
library, we are restricted to a magnitude limit of B=12, which precludes 
observing giant branches in clusters with this intermediate metallicity (e.g., 
M71).  As well, there is a danger in relying on stars selected from the metal-weak 
tail of a large photometric sample of nearby giants, since the tail of the 
distribution could be contaminated by stars with larger photometric errors.  
Second, coverage is sparse both for supergiants and for F and early G giants,
especially those with well-determined atmospheric parameters.  Third, there is
limited coverage at high metallicity, i.e., [Fe/H]$>$+0.2, especially given the
ongoing debate concerning the existence of super metal-rich stars (e.g., Taylor 
2002 and references therein).
Finally, we have only provided sparse coverage of O, B, and A stars.  The 
number of such stars with well-determined atmospheric parameters at low gravity
and below solar metallicity is very limited.  Furthermore, there is the problem
of emission contamination and peculiar chemical abundances in many of these 
stars.  In addition, some of the early-type stars may be rapid rotators,
thereby limiting the resultant spectral resolution.

\begin{figure}
\plotone{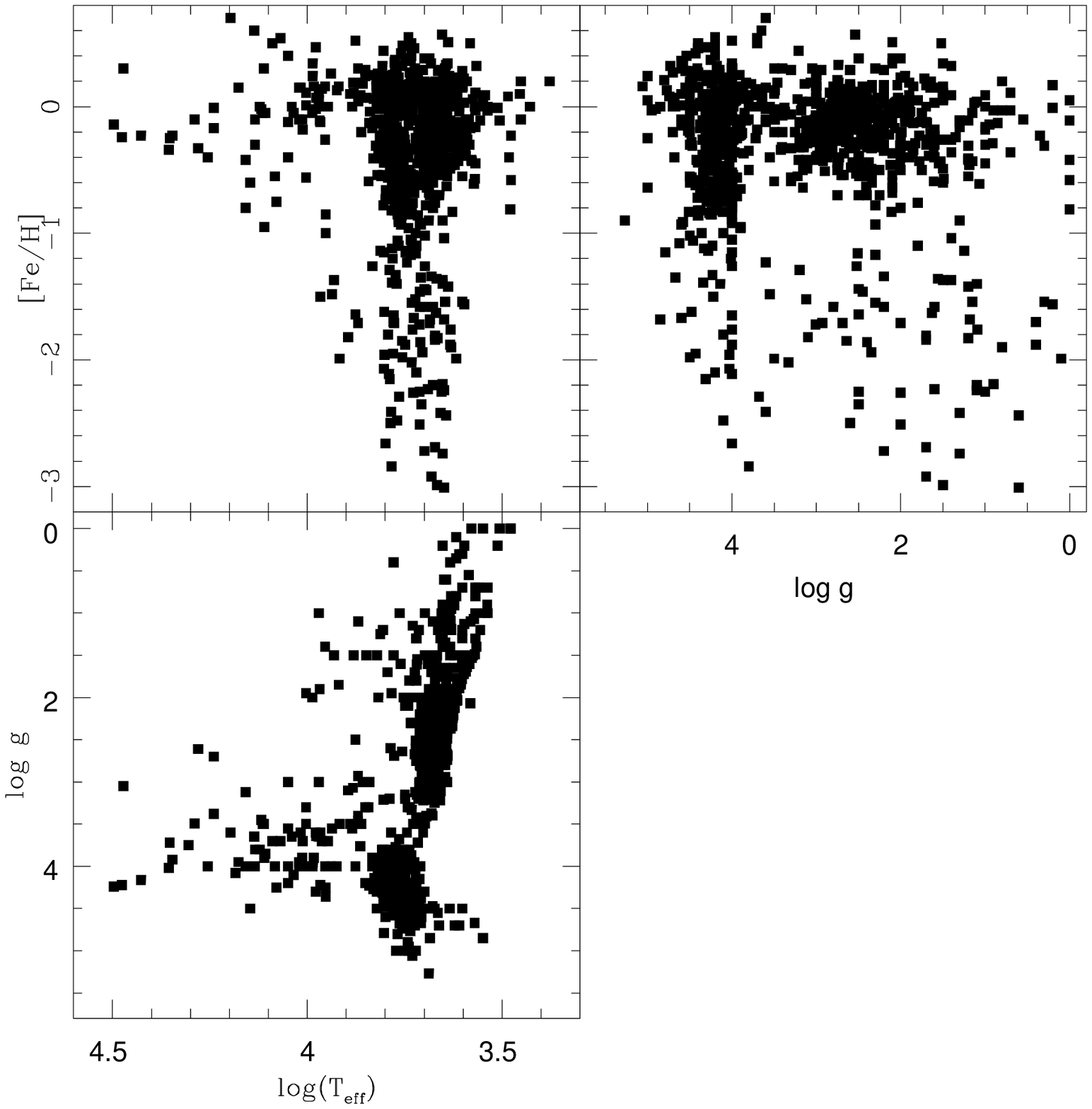}
\caption{The T$_{eff}$, [Fe/H], and log g coverage of the Coud\'e Feed spectral 
library. Coverage of the parameter space at temperatures hotter than 6000 K is
limited.  Note that a number of stars known to be dwarfs but with poorly 
determined log g have been assigned log g = 4.}
\label{fig:coude-pars}
\end{figure}

The coverage of atmospheric parameter space by our Coud\'e Feed spectral 
library is shown in Fig.~\ref{fig:coude-pars}.  As mentioned above, coverage is
generally quite extensive for T$_{eff}$ below $\sim$6300 K (log 
T$_{eff}$$<$3.8), but rapidly thins out at higher temperature, and at 
[Fe/H]$>$+0.2.

\section{Reductions and Calibrations}

The first stage of instrumental calibration is performed at the telescope.
All exposures are automatically corrected for amplifier bias levels using
the overscan region and then trimmed of the overscan and bad edge columns
and lines.  Sequences of bias exposures at the beginning of the night's
observations and quartz lamp flat fields at the beginning and end of
the night, typically 20 in a sequence, are obtained.  The bias sequence
is averaged into a master bias which is then automatically applied to
all subsequent calibration and stellar exposures.  The flat fields,
corrected by the master bias, are combined into master flat fields,
though these are not applied at the telescope.  A sequence of typically
five comparison lamp exposures are combined.  During the night the
stellar and comparison lamp exposures are automatically bias calibrated
and trimmed.  The individual calibration exposures and the raw stellar
exposures are not kept except in the observatory's failsafe tape archive.

The bias calibrated data, including the master flat fields and
arc comparisons, are processed by a pipeline built using the
IRAF\footnote{IRAF is distributed by the National Optical Astronomy
Observatories, which are operated by the Association of Universities
for Research in Astronomy, Inc., under cooperative agreement with the
National Science Foundation.} system and tasks.  At the highest level
the stages of the reductions are 2D flat field calibration, extraction
into 1D spectra,  dispersion calibration, telluric removal, heliocentric
and radial velocity corrections, continuum calibration, linear dispersion
resampling, and stacking and splicing of all exposures for a single star.
Documentation and header standardization is also performed along the way.

The 2D flat field calibration involves normalizing the master flat field
by a smooth curve approximately representing the quartz lamp spectrum.
Because we later apply a continuum calibration, the actual detailed shape of
the quartz lamp normalization is not critical.  The normalized flat field
provides calibration of the relative pixel responses, the slit function,
and the fringing which occurs in the redder wavelength exposures.

The 2D flat field is applied to each 2D stellar exposure.  Sometimes there
is only one master flat field, but when there are two, obtained at the
beginning and end of the night, the one nearest in time to the stellar
exposure is used.  This minimizes effects caused by flexure of the CCD
relative to the rest of the spectrograph during the night as the dewar
weight changes.

The extraction of the stellar spectra from the 2D long slit format to the
1D format is performed using the {\tt kpnoslit.doslit} IRAF task, in itself
a type of pipeline.  In this stage the position of the star in the image
is determined.  An extraction window is defined, centered on the peak of
the profile, with a width set to the full width of the profile at 5\% of
the maximum.  This adjusts automatically for seeing and guiding variations.
Two regions on either side of the profile window are used to define and
subtract a linear interpolation for the background in the extraction
window.

The extraction to a 1D spectrum consists of summing the background
subtracted pixel values along each column within the extraction window.
The primary sum is weighted by estimates of the variance in each pixel,
and pixel values that deviate by many sigma from the expected profile
determined from nearby wavelengths are rejected.  This is an implementation
of the "optimal variance-weighted algorithm" (see Horne 1986).  Since most of
the sources are fairly bright the main effect of this method is to greatly
reduce contamination from cosmic rays.  For reference, the full extracted
data set also includes the simple sum without weighting or rejection,
the background, and an estimate of the uncertainty in the extracted value
derived from the uncertainties in the summed pixels.

After extraction the nearest comparison lamp exposures are extracted over
the same pixels and a dispersion function is fit to the line positions
versus laboratory wavelengths.  The dispersion function assigned to the
stellar exposure is the average of the two weighted by the proximity
in time.

The telluric corrections are applied next.  Note that this is done prior
to radial velocity corrections because telluric features are fixed in
the observed frame of reference.  
For each program star a suitable telluric observation (same night
observation of a star from a list of telluric ``standards'') is selected; the
telluric standards are all rapidly rotating B stars. The
telluric star is continuum normalized with a cubic spline of 5 pieces fit to
non-telluric regions, where the telluric regions to be scaled and removed
are specified in Table~\ref{tab:telluric}. After the continuum fitting,
the non-telluric
regions are set to 1 so that that those regions do not modify the 
matching regions in the program star.  
The telluric features are removed by scaling and dividing the
template telluric regions into the stellar data.  The scaling and any
small wavelength shift are determined by minimization of the root mean
square in the corrected spectra over the telluric regions.
Hence, in effect, the telluric
regions in the telluric stars are continuum normalized from nearby
regions and only those regions are used to correct the program stars.
For the program stars, data outside the telluric regions is not modified by 
this procedure.

To enhance the ease of use of the spectral library, we have adjusted
all of the final spectra to heliocentric rest wavelength.  For all but 128
stars in the library, radial velocities are available from the literature,
and are taken from SIMBAD.  For the remaining 128 stars we use the {\tt
fxcor} routine in IRAF to determine radial velocities.  For 125 of these we use
the F8V star HD30562 as a radial velocity template.  For the remaining
3 stars, HD24398, HD24760, and HD87344, which are B stars, we use the
B star templates HD3360 and HD225132.  The radial velocities have been
obtained using the spectral region $\lambda$$\lambda$4000 - 4700 \AA.

As noted earlier, the combination of narrow slit, field rotation,
observations taken in non-photometric conditions, and limitations in
setting the slit to the parallactic angle makes flux calibration of our
data problematic.  Therefore, to smoothly join the different grating
settings taken over many years and to provide a continuum shape suitable
for automatic classifications algorithms, we fit each observation to
a spectral energy distribution standard with a close match in spectral type.
This requires a library of flux-calibrated spectra that covers both a large
range in spectral type and the 3400 - 9500 \AA \ wavelength
range of our spectra.

Unfortunately, there is no available library of spectrophotometry that
covers our extended wavelength region as well as providing full coverage
of the stellar atmospheric parameter space.  In fact, the existing
spectrophotometric libraries primarily cover MK spectral types and
luminosity classes, thus do not incorporate metal-poor stars.  We have
chosen the Pickles (1998) compilation as our spectrophotometric database.
Note that this library is normalized to unity at 5550~\AA \
and, hence, our library is also normalized to this point.  Note that this
normalization does not actually locate a ``true'' continuum at 5550~\AA,
since many faint spectral features at the location of the normalization 
point prevent the location of a true continuum there.

In order to handle the 14 cases for which there is no library spectral energy distribution
(SED), such as
for the C and S spectral types, or no known spectral type, such as some
of the fainter metal-poor stars, we added a dummy SED with unit continuum
everywhere.  This SED is in the same format as the rest of the library so
that the continuum matching is performed in the same way for all stars.

The continuum matching is carried out by integrating the SED standard and the
observation into adjacent 10~\AA \ bins.  Bins that cover strong spectral
features and/or telluric features are excluded.  (The complete set of
bins used 
may be found at {\em The Indo-U.S. Library of Coud\'e Feed Stellar
Spectra} home page at {\tt http://www.noao.edu/cflib/}.)
A smooth transformation 
function, specifically, a 10th order Chebyshev (with iterative rejection to
remove deviant points), is fit to the
ratio, expressed in magnitudes, of the observed and standard continuum
values.  The function is then applied to each pixel value in the
spectrum to produce the continuum calibrated versions of the data.

The Pickles continuum standard applied is identified in the header
information for the spectrum, as well as in the accompanying database
file, as is described in \S5.  For stars with spectral type earlier than
F0, our strategy is to find the closest Pickles star in spectral class,
since the Balmer discontinuity and spectral shape is more temperature
sensitive than gravity sensitive at the higher temperatures.  For stars
of type F0 and later we place first emphasis on attaining the correct
luminosity class.  As noted above, there are 14 stars with a flat continuum
applied.  Users of our library also need to be aware that the assignment
of a continuum standard for metal-poor stars is of limited validity.

To create a final spectrum from the separate exposures in different
grating settings, a two step process is followed.  First, the spectra
are interpolated onto a linear wavelength sampling, with a common start
wavelength and 0.4~\AA/pixel wavelength step.  Then the pieces are stitched
together into a single final spectrum.  In regions of wavelength overlap
between exposures, the overlap pixel values are averaged.  In cases where
gaps appear between grating settings, the gap pixels are set to 0.0001,
and the gap regions are recorded in the header and in the master database
file (see \S5).  Since the individual pieces are fit to the same Pickles
continuum SED, the overlap regions are automatically smoothly joined.

\section{The Spectral Library}

The completed spectral library consists of two principle components. The
first is the collection of spectra with pertinent information provided in
the headers.  The second is a database table that contains the relevant
information about each star.  The table allows for easy perusal of the
contents of the library.  Here we present a brief description of the
spectra and the database table.  The complete description of the products
may be found at {\em The Indo-U.S. Library of Coud\'e Feed Stellar
Spectra} home page at {\tt http://www.noao.edu/cflib/}.

\subsection{Spectra}

The library provides several versions of the spectra for each star to allow
the user to interpret the effects of various stages in the pipeline
reductions.  First the individual extractions from each observation are
available.  The pixel values are in detector counts with a gain of 2.3
electrons per count.  The extractions include two versions of the spectrum,
the variance weighted and cosmic ray cleaned version and the simple sum.
They also include the subtracted background and the estimated uncertainties
based on the photon statistics of the detector pixels.  No interpolation
between pixels has been applied and the wavelengths are defined by a
dispersion function and a wavelength vector.

The variance weighted spectra and uncertainties, again in the original
pixel wavelength bins, are provided with the telluric removal and
continuum calibration.

The multiple observations of a star are merged together in another
product.  The continuum calibrated spectra for each star are
interpolated to a uniform sampling of 3465~\AA \ to 9469~\AA \ in steps of
0.4~\AA.  As described earlier, a 0.0001 value is used when no observation
covered a particular wavelength and the average is used when multiple
exposures are available.  This version does not include uncertainty
information in the first release.

The available data formats are described in detail on the Library home page.
The formats include FITS binary tables, FITS images, and simple
text files.  The FITS files consist of multiple extensions to collect
all data from a single star in a single file.  The image format
uses an IRAF convention for spectral data.

The spectral files also contain header information from the original raw
data and information derived from Simbad and the database of stellar
parameters.  Most of this header information is self-explanatory, hence we
simply touch on a few key parameters here.  For the individual pieces of
the spectra, arising from specific grating settings, we include information
on the grating and its tilt.  For the stitched-together final spectrum, we
include a parameter, COVERAGE, which gives the start and end
wavelength of the spectrum.  While all spectra have been rebinned to the
same start and end wavelengths, if the coverage actually starts at a longer
wavelength than the nominal start wavelength, the first pixels are filled
with 0.0001.  The same applies for any pixels at the red end which are
beyond the extent of the actual data.  The COVERAGE parameter gives the
true start and end wavelengths of the data.  There is also a GAPS
parameter, which gives the wavelength intervals for which gaps exist, due
to underlap between grating settings or to missing grating settings.

\subsection{Database of Stellar Parameters}

Information on each star in the database is included in the header of the
spectral data files.  For ease of use, we have also summarized much of the
essential header information in both HTML and FITS Binary Table format at
the Library home page at {\tt http://www.noao.edu/cflib/}.  For printing
purposes, in this paper we have subdivided the master table into the three 
Tables~\ref{tab:parameters}, \ref{tab:parameters2}, and \ref{tab:parameters3}.  We print 
only the first 10 lines of each table, while the full tables appear in the
electronic version of the Journal and at the Library home page.
Key elements of Table~\ref{tab:parameters} are as
follows.  The first column contains the primary star ID, 
the second and third columns
contain the FK5 RA and Dec (2000.0) coordinates (as found in SIMBAD), and 
columns (4) and (5) contain the apparent V magnitude and B--V color
(as found in SIMBAD).
In column (6) is given the number of individual spectra obtained for the
star, column (7) provides the wavelength region covered by the spectrum, and
in column (8) we list any gaps present in the wavelength coverage.
In Table~\ref{tab:parameters2} the first column is again the primary star designation.
In columns (2) and (3) are listed the Spectral Type (from SIMBAD) and the
spectral type that we selected from the Pickles (1998) spectrophotometric 
library used for the continuum calibration.  If no spectral type match could
be made, a ``flat'' designation is given, to indicate that the continuum
has been normalized to unity.  Next is the radial velocity,
followed by the atmospheric parameters, T$_{eff}$, 
log g, and [Fe/H].  In column (8) is given the reference code for the
atmospheric parameters. 
Finally, in Table~\ref{tab:parameters3} we give the primary star ID in column (1),
and then follow with alternate name designations.

\subsection{Preliminary Artificial Neural Network Analysis}

As part of an effort to cross-check the reliability of both our reductions and
the assembled information on the nature of the stars, we have conducted a
preliminary automated analysis of the reduced spectra using an Artificial
Neural Network (ANN) scheme.  The ANN approach used here has been previously 
described in Gulati \etal (1994) and Singh, Gulati, \& Gupta (1998).  Our
goal was to compare the spectral type for each star that we selected from 
the Pickles (1998) library, to use for continuum-calibration, against the actual
listed MK spectral type in the SIMBAD database.  Any major discrepancy between
our selected spectral type and the derived ANN spectral type could indicate 
either a typographical error in our selection of the Pickles type, an error in 
the MK type given in SIMBAD (which we used as a basis for selecting the nearest
available Pickles type), or some problem in our reduction procedure for that
star.  

Since we are making a cross-check with MK spectral types, we therefore
restricted the ANN analysis to the blue spectral region of 
$\lambda$$\lambda$3510 - 4700 \AA.  For the analysis we used the 158 spectra
from the Jacoby, Hunter, \& Christian (1984; hereafter JHC) continuum-calibrated 
spectral library as the training set.  These spectra have a resolution of 
4.5 \AA \ FWHM, and we rebinned them to a sampling of 1.0 \AA/pixel.   We then 
broadened and rebinned our coud\'e feed spectra to the same resolution and 
sampling as the JHC spectra before applying the ANN analysis with the JHC
spectra as the training set.  We converted MK spectral types to a numerical code
using the following scheme.  The alphabetic MK class is coded into a numeric, 
with type O given a numeric code of 1000, B is 2000, F is 3000, etc.  The
decimal subclasses are multiplied by 100 and added in, thus a G2 star is coded
as 5200.  The luminosity class is coded as follows: types I, II, II, IV, and V
are coded as 1.5, 3.5, 5.5, 7.5, and 9.5, respectively.  The luminosity class is
also added in, hence a G2V is coded as 5209.5.  Evidently, the luminosity class
is given very low weight in the analysis.  It was found that after application 
of the ANN 
scheme, we located approximately a dozen stars which had been inadvertently
assigned the wrong Pickles spectral class.  These stars were accordingly
reassigned the correct nearest Pickles class and reanalyzed during the final
reduction stage.  Overall, we find a 2-$\sigma$ scatter between the assigned
Pickles type and the ANN-derived type of only 1.1 decimal subclasses.  A few
outliers, other than the ones with wrongly listed Pickles types, have been 
found, some of them being low metallicity stars, for which MK spectral types
are rather meaningless.  A few other cases involve late M stars, for which the
blue spectral region is quite problematic for applying spectral classification.
The remainder involve confusion between B stars with similar Balmer line
strengths as A stars.  We reserve a more complete analysis with ANN techniques
to a future paper.

\subsection{Availability of Spectral Database}

As mentioned above, the entire spectral library is publicly available from
the Library home page at {\tt http://www.noao.edu/cflib/}.  FTP access is
also provided from \\
{\tt ftp://ftp.noao.edu/cflib/}.  This currently
provides version 1.0 of the library.  We expect to release future versions,
with improved calibrations and uncertainties and with updated stellar
atmospheric parameters.  The library will eventually be incorporated in the
NOAO Science Archive and the Virtual Observatory to provide search and
browse capabilities.  We encourage interested users to advise us of any
defects in the spectral library, and to alert us to improved atmospheric
parameter information for stars, as well as to any new spectrophotometry
that we may be able to use to improve continuum calibration on some or all
of the stars.

\section{Examples of Spectra}

The primary purpose of this paper is to describe the spectral database in order
to make it as generally useful as possible to a variety of investigators. 
However, in the following Section we show a few examples of the spectra,
both to illustrate the quantity of information contained in the
spectra, and also to illustrate some of the challenges involved in developing
a reliable automated classification for stellar spectra.

\begin{figure}
\plotone{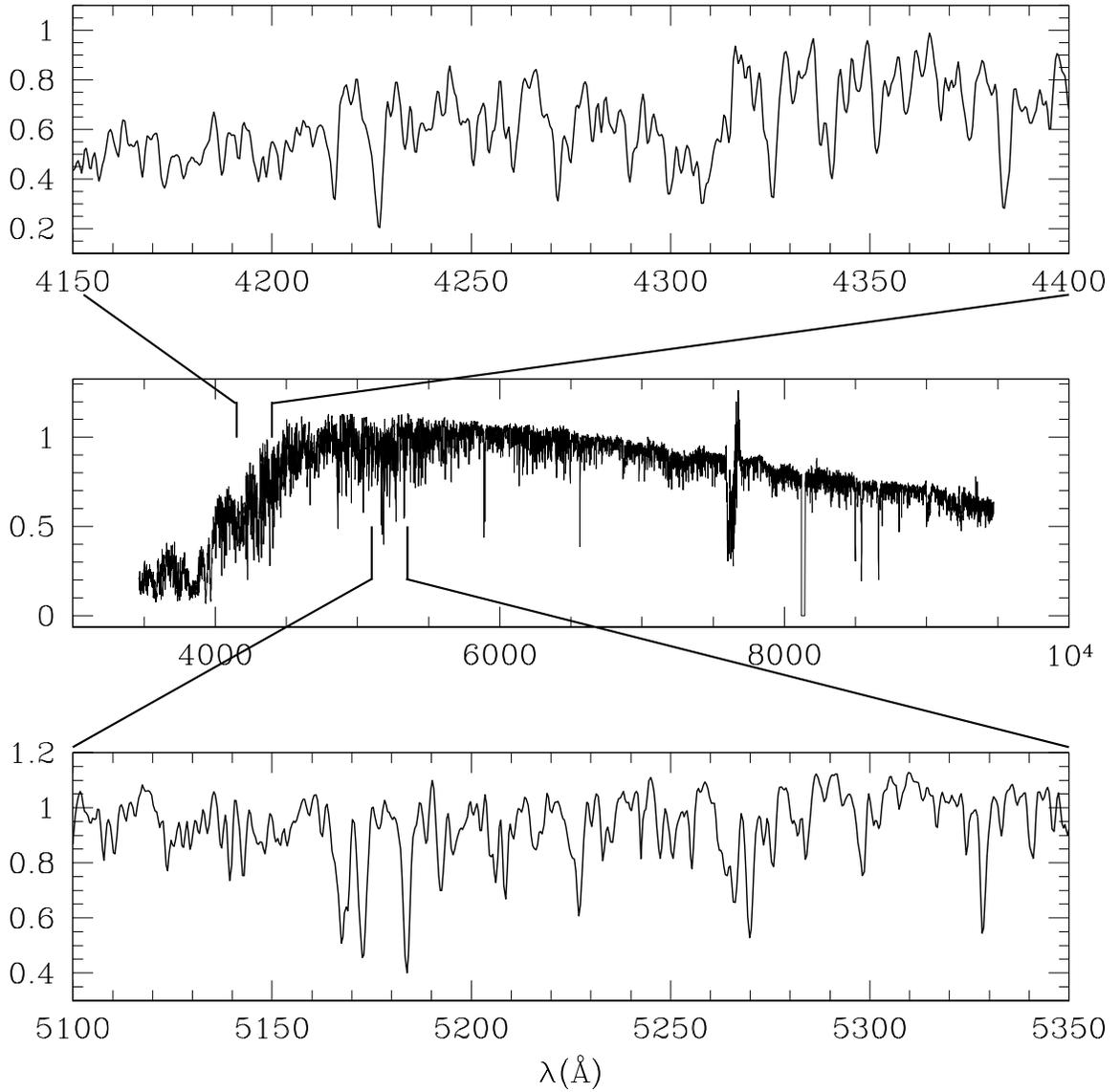}
\caption{The spectrum of the K0III HD4128 is plotted, with expanded regions
to show the large information content of the spectrum. Given that the S/N ratio
exceeds 100:1 per pixel, essentially all features discernible in the spectrum 
are real.}
\label{fig:allspec}
\end{figure}

\begin{figure}
\plotone{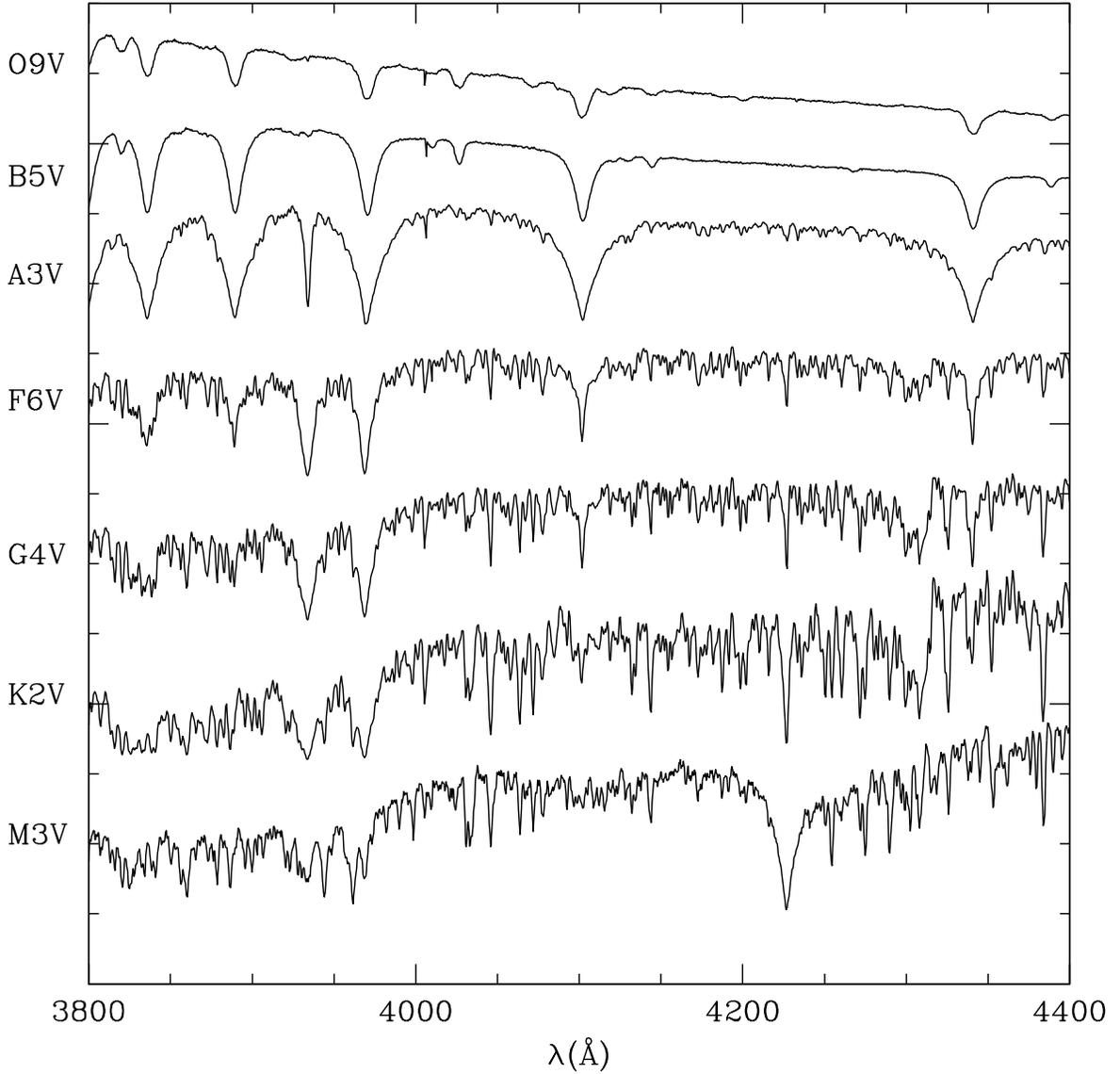}
\caption{Spectra of 7 dwarf stars, covering a large range in MK spectral type,
are plotted in the blue, to illustrate the basic dependence of spectral
features with MK type.  The stars plotted, top to bottom, are HD149757,
HD158148, HD163624, HD173667, HD52711, HD166620, and HD173739.  The spectral
types are listed on the vertical axis.}
\label{fig:MK}
\end{figure}

In Fig.~\ref{fig:allspec} we show the entire 3400 - 9500 \AA \ spectral region
covered by our observations for the K0III star HD4128.  The point of the
plot is to show the large amount of information present in the spectrum,
which is evident in the expanded wavelength regions.  In Fig.~\ref{fig:MK}
a series of 7 dwarf stars covering the range of spectral types from O through M are
plotted, to illustrate the basic behavior of the spectra with changing
temperature.  We have restricted the plots to only the blue spectral region,
in the wavelength interval where much of MK classification has been carried
out.

\begin{figure}
\plotone{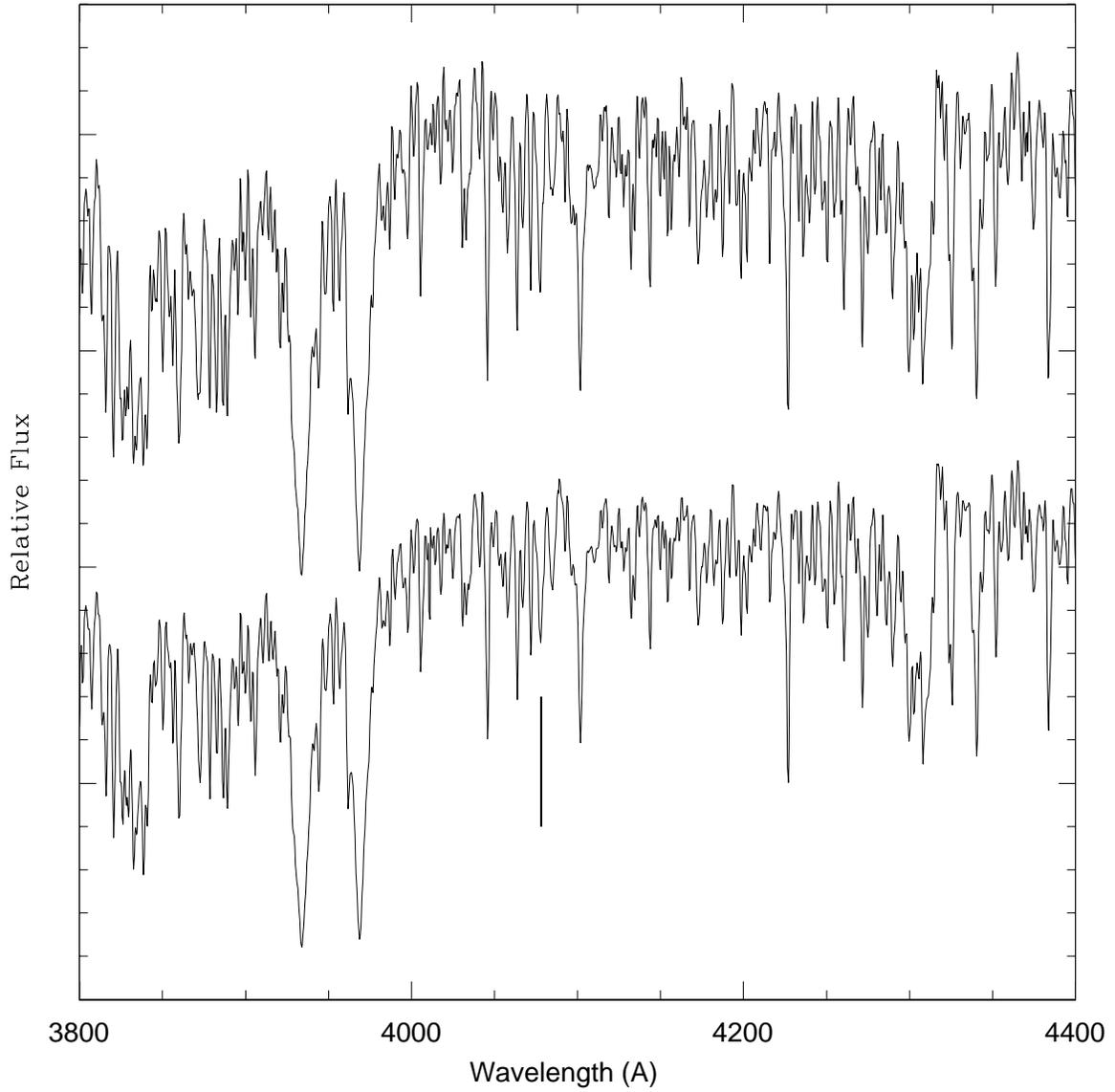}
\caption{The spectrum of the metal-rich star HD30562 (top spectrum), with 
[Fe/H]=+0.14, is compared with that of the metal-poor star HD157214 (bottom
spectrum), with [Fe/H]=-0.41.  The remarkable similarity of the two spectra
illustrates the degenerate effects of T$_{eff}$ and [Fe/H] on the appearance
of the spectra.}
\label{fig:FeH}
\end{figure}

In Fig~\ref{fig:FeH} we compare the spectra of two stars which have similar
spectral type, but very different T$_{eff}$ and [Fe/H].  Specifically, the
upper spectrum is of the star HD30562, which has T$_{eff}$=5860 and [Fe/H]=+0.13,
while the lower spectrum is of HD157214, which has T$_{eff}$=5600 and 
[Fe/H]=-0.58.  The factor of four difference in metallicity is so well
balanced by the hotter temperature in the more metal-rich star that the two
spectra are virtually indistinguishable.  Essentially, Fig.~\ref{fig:FeH}
illustrates the long-known degeneracy between T$_{eff}$ and [Fe/H], which for
decades obscured the presence of a metallicity spread and chemical enrichment
history in the Galaxy, and currently plagues the disentanglement of age from 
metallicity effects in the integrated spectra of galaxies (e.g., Worthey 1994).

\begin{figure}
\plotone{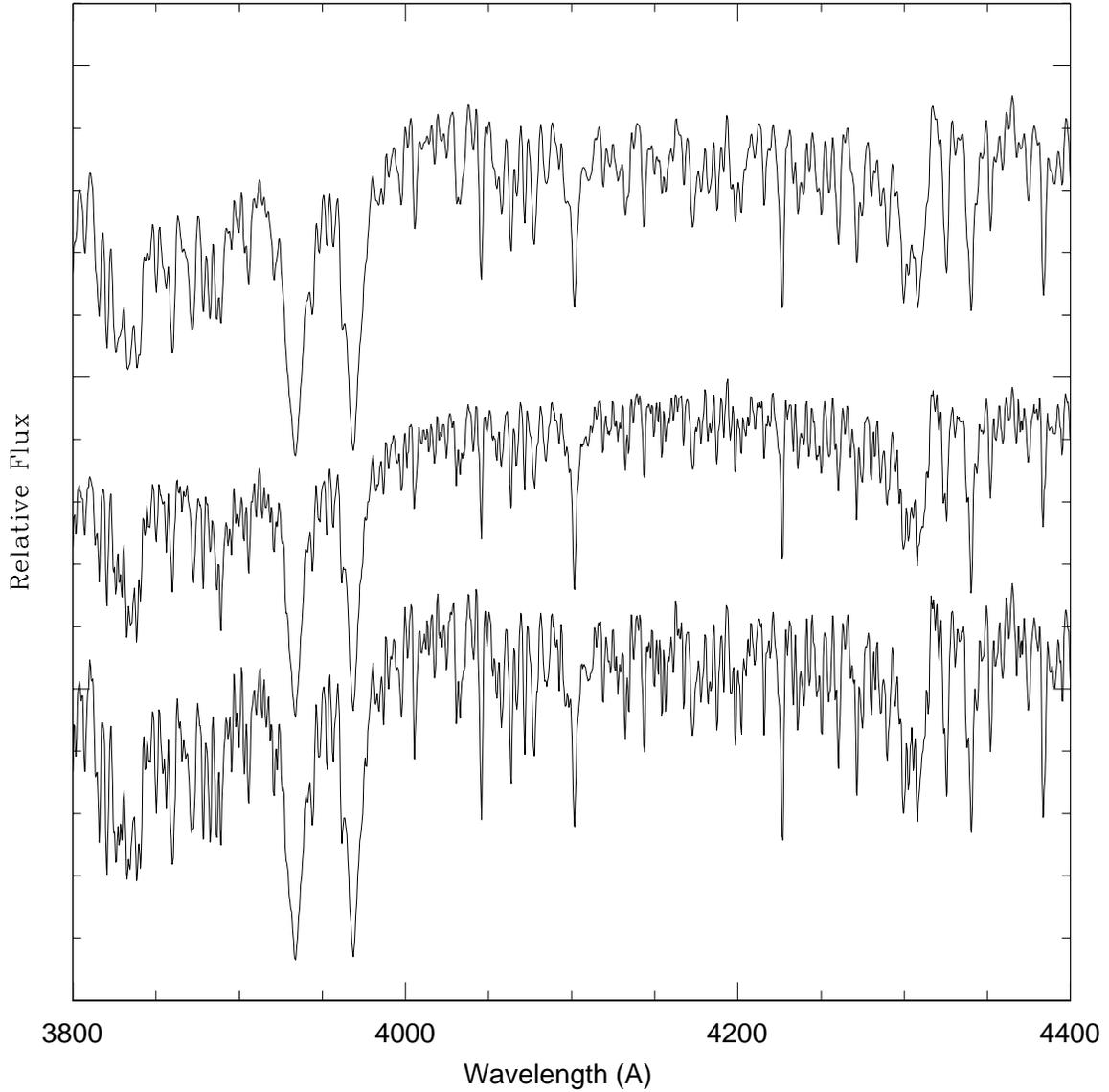}
\caption{The unsmoothed (bottom) and smoothed (top) spectra of HD30562 are
compared to the unsmoothed spectrum of the metal-poor star HD105755 (middle).
In the top spectrum HD30562 has been smoothed with a gaussian $\sigma$ of
1.5 pixels.}
\label{fig:smooth}
\end{figure}

There are several aspects to the Figure worth emphasizing.  First is the 
remarkably strong degeneracy of temperature and metallicity effects.  The
second is the difficulty in measuring any one feature in the blue at 
intermediate spectral resolution, for anything but the hottest stars.  To
further illustrate the latter point, in Fig~\ref{fig:smooth} we show the 
spectrum of HD30562 at its original spectral resolution (bottom), and 
smoothed with a $\sigma$ of 1.5 pixels (top), bracketing the spectrum of the
metal-poor ([Fe/H]=-1.2) dwarf HD105755 (middle).  As can be seen in the
Figure, a loss of spectral resolution gives the strong impression of a
weaker lined spectrum, in that the smoothed spectrum of HD30562 on top comes
considerably closer to duplicating that of HD105755 in the middle than the
unsmoothed spectrum of HD30562.
This effect illustrates the potential degeneracy between
spectral resolution and line strength, a confusion which can also be 
challenging to disentangle.  

A further point is that the great majority of
information in the spectra is highly redundant, with only a handful of key
features supplying unique information.  For example, a close look at the
strength of the SrII$\lambda$4077 feature in Fig.~\ref{fig:FeH} reveals that it
is noticeably 
stronger in HD30562 than in HD157214.  This is due to the lower surface
gravity in HD30562 (log g=3.75) than in HD157214 (log g=4.27), which is 
reflected in the ionization balance that affects the strength of the singly
ionized strontium feature relative to neighboring neutral iron lines.  Thus
the clean separation of atmospheric parameters requires a careful 
optimization of the spectral features to be employed, which poses a considerable
challenge to automated spectral classification methods.

\begin{figure}
\plotone{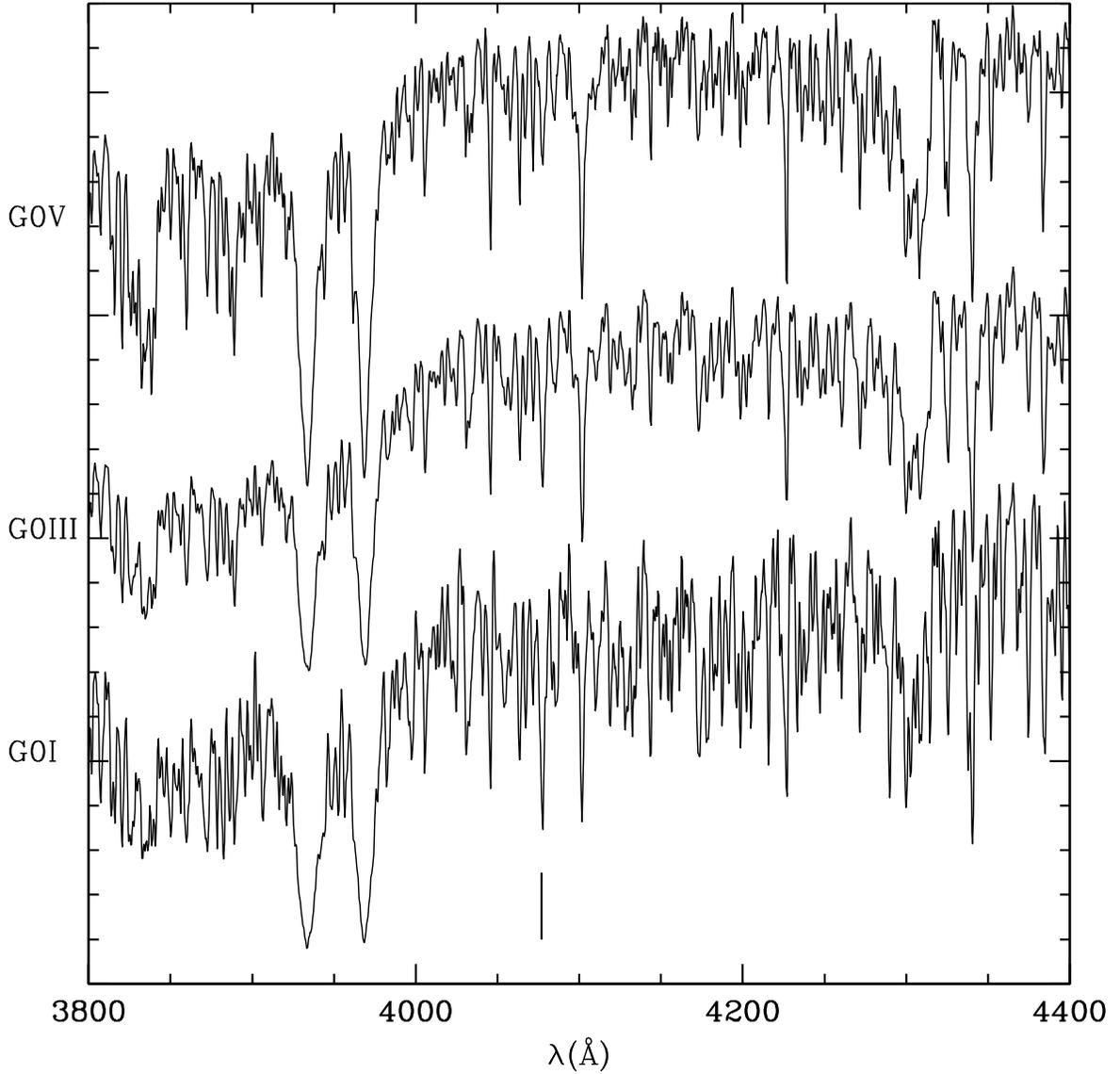}
\caption{The spectra of three G0 stars with differing luminosity classes are
compared, again in the restricted blue wavelength region.  From top to bottom
they are the G0V star HD184499, the G0III star HD111812, and the G0Ib star HD204867.  The 
singly ionized SrII$\lambda$4077 line, which shows strong sensitivity to
luminosity class (i.e., surface gravity), is specifically marked in HD204867.}
\label{fig:grav}
\end{figure}

To further illustrate the subtlety of atmospheric effects, in Fig~\ref{fig:grav}
we plot the spectra of three late F/early G stars covering a range in surface
gravity.  The top star, HD187691 has log g=4.4, the middle star, HD111812 is
a luminosity class III G0 star, and the bottom spectrum is of the star
HD204867, a G0Ib star with log g=1.3.  The most striking change along the
sequence is in the strength of the SrII$\lambda$4077 feature.

In short, the above illustrations indicate that optimization of the specific
spectral features and criteria to be employed in extracting atmospheric
parameters from stellar spectra, as well as the methods used to define
the features, requires careful consideration. It is hoped that this
library will serve as a useful tool in that regard for a variety of studies
that will advance our understanding and utilization of stellar spectra.

\acknowledgements

We wish to thank Lindsay Bartholomew, Ravi Gulati,
Lauren Johnson, Lewis Jones, David Moeschler, Jane 
Moran, Melissa Nysewander, Christina Reynolds, Jesse Richuso, and Celeste 
Yeates, all of whom made observations at the Coud\'e Feed for this project.  Their
participation in the coud\'e feed observations has made the completion of this
project possible.
Daryl Willmarth's expertise, dedication, and timely support of the coud\'e feed 
telescope and spectrograph over the years has been invaluable for this project,
and is gratefully acknowledged here. 
We also thank Bill Binkert for setups and support during
the first half nights of several observing runs. 
The comments of the referee, Claus Leitherer, led to many 
improvements in the text of the paper.
This research was partially supported 
by a joint NSF/DST grant to the University of North Carolina and to the 
Inter-University Center for Astronomy and Astrophysics in Pune, India.  The
research has made use of the SIMBAD database, operated at CDS, Strasbourg, 
France.

%\appendix
%
%\section{Emission Corrections}
%
%The detection of an emission spectrum superimposed on an absorption spectrum is
%
%\clearpage
%

\newpage

\clearpage

\newpage

\pagestyle{empty}

\begin{deluxetable}{ll}
\tablenum{1}
\tablecolumns{2}
\tablewidth{0pt}
\tablecaption{Wavelength Regions for Telluric Lines. \label{tab:telluric}}
\tablehead{
\colhead{Start Wavelength} & \colhead{End Wavelength} } 
\startdata
6863 & 6964\\
6983 & 7058\\
7163 & 7400\\
8128 & 8393\\
8902 & 9600\\
\enddata
\end{deluxetable}
\label{tab:telluric}

\pagestyle{empty}
\newcommand\cola {\null}
\newcommand\colb {&}
\newcommand\colc {&}
\newcommand\cold {&}
\newcommand\cole {&}
\newcommand\colf {&}
\newcommand\colg {&}
\newcommand\colh {&}
\newcommand\eol{\\}
\newcommand\extline{&&&&&&&\eol}

\begin{deluxetable}{llrrrrll}
\tablenum{2}
\tablecolumns{8}
\footnotesize
\tablecaption{Observational Parameters\label{tab:parameters}}
\tablewidth{0pt}
\tablehead{
\colhead{\bf Star ID} &
\colhead{\bf \hspace{0.7ex}$\alpha$ (2000)} &
\colhead{\bf $\delta$ (2000)\hspace{0.6ex}} &
\colhead{\bf V\hspace{1.4ex}} &
\colhead{\bf B-V} &
\colhead{\bf N} &
\colhead{\bf \hspace{-0.5ex}Coverage (\AA)} &
\colhead{\bf Gaps (\AA)}
}

%\multicolumn{8}{c}{\large\bf Table 2: Observational Parameters}\\
%\multicolumn{8}{c}{\null}\\

%\cola {\bf Star ID} 
%\colb {\bf \hspace{0.7ex}$\alpha$ (2000)} 
%\colc {\bf $\delta$ (2000)\hspace{0.6ex}} 
%\cold {\bf V\hspace{1.4ex}}
%\cole {\bf B-V}
%\colf {\bf N}
%\colg {\bf \hspace{-0.5ex}Coverage (\AA)}
%\colh {\bf Gaps (\AA)} 
%\eol
%\hline\\
%\endfirsthead

%\multicolumn{8}{c}{\bf Table 2 (continued)}\\
%\multicolumn{8}{c}{\null}\\

%\cola {\bf Star ID} 
%\colb {\bf \hspace{0.7ex}$\alpha$ (2000)} 
%\colc {\bf $\delta$ (2000)\hspace{0.6ex}} 
%\cold {\bf V\hspace{1.4ex}}
%\cole {\bf B-V}
%\colf {\bf N}
%\colg {\bf \hspace{-0.5ex}Coverage (\AA)}
%\colh {\bf Gaps (\AA)} 

%\eol
%\hline\\
%\endhead

%\\
%\hline\\
%\endfoot
\startdata
\cola G 4-44\colb 02:51:58.36\colc +11:22:11.9\cold  8.38\cole  0.54\colf  3\colg 3465.0-7265.8\colh \eol
\cola G 5-40\colb 03:27:39.70\colc +21:02:30.0\cold 10.79\cole  0.57\colf  3\colg 3465.0-7269.4\colh \eol
\cola G 7-6\colb 03:50:22.97\colc +17:28:34.9\cold  7.52\cole  0.66\colf  4\colg 3465.0-8126.2\colh \eol
\cola G 11-45\colb 12:12:28.84\colc -03:05:04.1\cold  7.50\cole  0.65\colf  4\colg 3465.0-8127.4\colh \eol
\cola G 12-21\colb 12:12:01.37\colc +13:15:40.6\cold 10.18\cole  0.47\colf  4\colg 3465.0-8104.2\colh 5413.0-5465.4\eol
\cola G 12-22\colb 12:12:57.53\colc +10:02:15.8\cold  7.92\cole  0.79\colf  3\colg 3465.0-8106.2\colh 4763.8-5467.4\eol
\cola G 12-24\colb 12:13:13.12\colc +10:49:18.0\cold  7.56\cole  0.68\colf  6\colg 3465.0-9451.8\colh \eol
\cola G 15-20\colb 15:22:42.55\colc +01:25:07.1\cold  8.30\cole  1.00\colf  4\colg 3465.0-8121.0\colh \eol
\cola G 16-13\colb 15:50:58.93\colc +08:25:23.8\cold 10.01\cole  0.59\colf  6\colg 3465.0-9452.6\colh 5415.8-5462.2\eol
\cola G 16-32\colb 16:09:11.21\colc +06:22:43.3\cold  5.94\cole  1.03\colf  6\colg 3465.0-9469.0\colh 5415.0-5461.8\eol
\enddata
\tablecomments{Table 2 is published in its entirety in the electronic
edition of the Astrophysical Journal Supplements.  A portion is shown here
for guidance regarding its form and content.}

\end{deluxetable}

%\label{tab:parameters}
%\label{tab:parameters}

\pagestyle{empty}

\begin{deluxetable}{lllrrrrl}
\tablenum{3}
\tablecolumns{8}
\footnotesize
\tablecaption{Physical Parameters\label{tab:parameters2}}
\tablewidth{0pt}
%\multicolumn{8}{c}{\large\bf Table 3: Physical Parameters}\\
%\multicolumn{8}{c}{\null}\\
\tablehead{
\colhead {\small\bf Star ID} &
\colhead {\small\bf Type} &
\colhead {\small\bf Pickles\hspace{-2ex}} &
\colhead {\small\bf V (km/s)\hspace{-2ex}} &
\colhead {\small\bf T${_{\mathrm{eff}}}$(K)\hspace{-1.5ex}}  &
\colhead {\small\bf log${_{10}}(g)$\hspace{-2ex}}  &
\colhead {\small\bf [Fe/H]\hspace{-1.2ex}}  &
\colhead {\small\bf Reference} 
}
%\eol
%\hline\\
%\endfirsthead

%\multicolumn{8}{c}{\bf Table 3 (continued)}\\
%\multicolumn{8}{c}{\null}\\
%
%\cola {\small\bf Star ID} 
%\colb {\small\bf Type} 
%\colc {\small\bf Pickles\hspace{-2ex}}
%\cold {\small\bf V (km/s)\hspace{-2ex}}
%\cole {\small\bf T${_{\mathrm{eff}}}$(K)\hspace{-1.5ex}} 
%\colf {\small\bf log${_{10}}(g)$\hspace{-2ex}} 
%\colg {\small\bf [Fe/H]\hspace{-1.2ex}} 
%\colh {\small\bf Reference} 
%\eol
%\hline\\
%\endhead

%\\
%\hline\\
%\endfoot
\startdata
\cola\small G 4-44\colb\small G5\colc\small G5V\cold\small    6.2\cole\small   5750\colf\small   4.11\colg\small  -0.69\colh\small 2000A\&A...353..722N (Nissen)\eol
\cola\small G 5-40\colb\small G0\colc\small G0V\cold\small -117.8\cole\small   5863\colf\small   4.24\colg\small  -0.83\colh\small 1997A\&A...326..751N (Nissen)\eol
\cola\small G 7-6\colb\small G0\colc\small G0V\cold\small   -9.3\cole\small   5594\colf\small   4.50\colg\small   0.07\colh\small 2001A\&A...369.1048P (Prugniel)\eol
\cola\small G 11-45\colb\small G4V\colc\small G5V\cold\small   14.1\cole\small   5725\colf\small  -~~~\colg\small   0.15\colh\small 1997A\&A...323..809F (Favata)\eol
\cola\small G 12-21\colb\small F2\colc\small F2V\cold\small   95.0\cole\small   5939\colf\small   4.23\colg\small  -1.33\colh\small 2000A\&A...353..722N (Nissen)\eol
\cola\small G 12-22\colb\small G8V\colc\small G8V\cold\small   -8.8\cole\small   5437\colf\small   4.77\colg\small   0.13\colh\small 1998A\&AS..129..237F (Feltzing)\eol
\cola\small G 12-24\colb\small G3V\colc\small G2V\cold\small  -29.5\cole\small   5337\colf\small   4.00\colg\small  -0.54\colh\small 1994AJ....107.2240C (Carney)\eol
\cola\small G 15-20\colb\small K3V\colc\small K3V\cold\small  -30.4\cole\small   4765\colf\small   4.47\colg\small   0.19\colh\small 1998A\&AS..129..237F (Feltzing)\eol
\cola\small G 16-13\colb\small G0\colc\small G0V\cold\small  -51.5\cole\small   5593\colf\small   4.00\colg\small  -1.15\colh\small 1994AJ....107.2240C (Carney)\eol
\cola\small G 16-32\colb\small K1.5IV\colc\small K1IV\cold\small   -4.0\cole\small   4768\colf\small   3.40\colg\small  -0.07\colh\small 1999A\&A...348..487R (Randich)\eol
%
%\end{longtable}
%
\enddata
\tablecomments{Table 3 is published in its entirety in the electronic
edition of the Astrophysical Journal Supplements.  A portion is shown here
for guidance regarding its form and content.}
\end{deluxetable}
\label{tab:parameters2}

\pagestyle{empty}
\def\extline{\eol}

\begin{deluxetable}{lllllll}
\tablenum{4}
\tablecolumns{7}
\footnotesize
\tablecaption{Cross-references for Star Names\label{tab:parameters3}}
%\multicolumn{7}{c}{\large\bf Table 4: Cross-references for Star Names}\\
%\multicolumn{7}{c}{\null}\\

\tablehead{
\colhead {\bf G} &
\colhead {\bf HD} &
\colhead {\bf BD} &
\colhead {\bf SAO} &
\colhead {\bf HR} &
\colhead {\bf HIP} &
\colhead {\bf Alternate}
}

%  {\bf G}
%& {\bf HD} 
%& {\bf BD} 
%& {\bf SAO} 
%& {\bf HR} 
%& {\bf HIP} 
%& {\bf Alternate} 
%\eol
%\hline\\
%\endfirsthead

%\multicolumn{7}{c}{\bf Table 4 (continued)}\\
%\multicolumn{7}{c}{\null}\\

%  {\bf G}
%& {\bf HD} 
%& {\bf BD} 
%& {\bf SAO} 
%& {\bf HR} 
%& {\bf HIP} 
%& {\bf Alternate} 
%\eol
%\hline\\
%\endhead

%\\
%\hline\\
%\endfoot
\startdata
\small G 4-44 & \small HD 17820 & \small BD+10 380 & \small SAO 93151 & \small  & \small HIP 13366 & \small G 76-35 \eol
\small G 5-40 & \small  & \small BD+20 571 & \small  & \small  & \small  & \small G 6-11 \eol
\small G 7-6 & \small HD 24040 & \small BD+17 638 & \small SAO 93630 & \small  & \small HIP 17960 & \small  \eol
\small G 11-45 & \small HD 106116 & \small BD-02 3481 & \small SAO 138647 & \small  & \small HIP 59532 & \small G 13-19 \eol
\small G 12-21 & \small HD 106038 & \small BD+14 2481 & \small SAO 99984 & \small  & \small HIP 59490 & \small G 57-39 \eol
\small G 12-22 & \small HD 106156 & \small BD+10 2391 & \small SAO 99991 & \small  & \small HIP 59572 & \small G 57-40 \eol
\small G 12-24 & \small HD 106210 & \small BD+11 2439 & \small SAO 99995 & \small  & \small HIP 59589 & \small G 57-41 \eol
\small G 15-20 & \small HD 136834 & \small BD+01 3071 & \small SAO 120966 & \small  & \small HIP 75266 & \small  \eol
\small G 16-13 & \small  & \small BD+08 3095 & \small  & \small  & \small HIP 77637 & \small  \eol
\small G 16-32 & \small HD 145148 & \small BD+06 3169 & \small SAO 121392 & \small HR 6014 & \small HIP 79137 & \small  \eol
%
%\end{longtable}
\enddata
\tablecomments{Table 4 is published in its entirety in the electronic
edition of the Astrophysical Journal Supplements.  A portion is shown here
for guidance regarding its form and content.}
\end{deluxetable}
%\label{tab:parameters3}

\end{document}